\documentclass[final]{svjour3}
\usepackage{graphicx}

\usepackage{url}
\usepackage{rotating}
\usepackage{amssymb}
\usepackage{amsmath}
\usepackage{multirow}
\usepackage{mathptmx}
\usepackage{kotex}
\usepackage[numbers]{natbib}
\usepackage{gensymb}
\usepackage{caption}
\usepackage{subcaption}
\makeatletter
\journalname{Journal of Low Temperature Physics}

\begin{document}

\newcommand{\hdblarrow}{H\makebox[0.9ex][l]{$\downdownarrows$}-}
\title{GroundBIRD : A CMB polarization experiment with MKID arrays}

\author{
    K. Lee$^{1}$ \and
    J. Choi$^{2}$ \and
    R. T. Génova-Santos$^{3,4}$ \and
    M. Hattori$^{5}$ \and
    M. Hazumi$^{6,7}$ \and
    S. Honda$^{8}$ \and
    T. Ikemitsu$^{8}$ \and
    H. Ishida$^{5,9}$ \and
    H. Ishitsuka$^{7}$ \and
    Y. Jo$^{1}$ \and
    K. Karatsu$^{10}$ \and
    K. Kiuchi$^{11}$ \and
    J. Komine$^{8}$ \and
    R. Koyano$^{12}$ \and
    H. Kutsuma$^{5,9}$ \and
    S. Mima$^{9}$ \and
    M. Minowa$^{11}$ \and
    J. Moon$^{1}$ \and
    M. Nagai$^{13}$ \and
    T. Nagasaki$^{9}$ \and
    M. Naruse$^{12}$ \and
    S. Oguri$^{9}$ \and
    C. Otani$^{9}$ \and
    M. Peel$^{3,4}$ \and
    R. Rebolo$^{3,4}$ \and
    J. A. Rubiño-Martín$^{3,4}$ \and
    Y. Sekimoto$^{14}$ \and
    J. Suzuki$^{8}$ \and
    T. Taino$^{12}$ \and
    O. Tajima$^{8}$ \and
    N. Tomita$^{11}$ \and
    T. Uchida$^{6,7}$	\and
    E. Won$^{1}$ \and
    M. Yoshida$^{6,7}$
}


\institute{
$^1$Korea University, 145 Anam-ro, Seongbuk-gu, Seoul, 02841, Republic of Korea\\ 
$^2$Institute of Basic Science, 70, Yuseong-daero 1689-gil, Yuseong-gu, Daejeon, Republic of Korea \\
$^3$Instituto de Astrof\'{i}sica de Canarias, E-38205 La Laguna, Tenerife, Spain \\
$^4$Departamento de Astrof\'{i}sica, Universidad de La Laguna, E-38206 La Laguna, Tenerife, Spain \\
$^5$Tohoku University, 2-1-1 Katahira, Aoba-ku, Sendai, Miyagi 980-8577, Japan \\
$^6$The High Energy Accelerator Research Organization (KEK), 1-1 Oho, Tsukuba, Ibaraki 305-0801, Japan \\ 
$^7$The Graduate University for Advanced Studies (SOKENDAI), Shonan Village, Hayama, Kanagawa 240-0193, Japan \\
$^8$Kyoto University, Yoshidahonmachi, Sakyo Ward, Kyoto, 606-8501 Japan \\
$^9$Institute of Physical and Chemical Research, 2-1 Hirasawa, Wako, Saitama 351-0198, Japan \\ 
$^{10}$Delft University of Technology, Mekelweg 5, 2628 CD Delft, Netherland \\
$^{11}$The University of Tokyo, 7-3-1 Hongo, Bunkyo-ku, Tokyo 113-8654, Japan \\
$^{12}$Saitama University, 255 Shimo-Okubo, Sakura-ku, Saitama 338-8570, Japan \\
$^{13}$National Astronomical Observatory of Japan, 2-21-1 Osawa, Mitaka, Tokyo 181-8588, Japan \\
$^{14}$Japan Aerospace Exploration Agency (JAXA), 7-44-1 Jindaiji, Higashi-machi, Chofu-shi, Tokyo 182-8522, Japan \\
\email{kmlee@hep.korea.ac.kr} 
}

\maketitle

\begin{abstract}  

GroundBIRD is a ground-based experiment for the precise observation of the polarization of the cosmic microwave background (CMB). To achieve high sensitivity at large angular scale, we adopt three features in this experiment: fast rotation scanning, microwave kinetic inductance detector (MKID) and cold optics. The rotation scanning strategy has the advantage to suppress $1/f$ noise. It also provides a large sky coverage of 40\%, which corresponds to the large angular scales of $l \sim 6$. This allows us to constrain the tensor-to-scalar ratio by using low $l$ B-mode spectrum. The focal plane consists of 7 MKID arrays for two target frequencies, 145 GHz and 220 GHz band. There are 161 pixels in total, of which 138 are for 144 GHz and 23 are for 220 GHz. This array is currently under development and the prototype will soon be evaluated in telescope. The GroundBIRD telescope will observe the CMB at the Teide observatory. The telescope was moved from Japan to Tenerife and is now under test. We present the status and plan of the GroundBIRD experiment.

\end{abstract}

\section{Introduction} 

The cosmic microwave background (CMB) is a isotropic blackbody radiation at the temperature of 2.7255 K \cite{WMAP}. The CMB has small anisotropies at the $10^{-5}~\mathrm{K}$ level in temperature and sub-$\mu\mathrm{K}$ level in polarization. The CMB anisotropies can be characterized by the angular power spectrum, and it provides constraints on numerous cosmological parameters \cite{Dodelson}.

The E-modes of CMB polarization have been well observed by various experiments \cite{WMAP} and lensing B-modes have been measured by other experiments, PolarBEAR and BICEP2 \cite{Polarbear, Bicep2}. The B-modes larger than degree scale, generated by primordial gravitational waves predicted by the inflation model, has not been detected yet \cite{Guth1981}.
The observation of B-modes from the primordial gravitational waves will reveal information about the very early universe hidden behind the background radiations.

GroundBIRD is aiming to measure the B-modes of CMB polarization \cite{Tajima2012}. The observation site is Teide observatory. Earlier this year, we transported the GroundBIRD telescope from Japan to Instituto de Astrof\'{i}sica de Canarias (IAC). In IAC, the telescope is assembled and is now under test. We are preparing for installation of the telescope at the Teide observatory in parallel. 
Once the assembly and tests are finished at IAC, we will move the telescope to the Teide observatory and continue the commissioning of the telescope with sky signals. 


\section{Overview of GroundBIRD}

The GroundBIRD telescope is designed for the precise observation of CMB polarization. The target sensitivity is to constrain the tensor-to-scalar ratio $r$ with a precision of $\sigma_r \sim 0.01$ with three year observation \cite{Oguri2016}.
To achieve high enough sensitivity to observe B-modes for large sky coverage, we employ three features: fast rotation scanning, microwave kinetic inductance detector (MKID) and cold optics \cite{Tajima2012}.

Aside from the thermal noise, the main source of systematics in CMB observations at large angular scale is $1/f$ noise due to the atmospheric fluctuations. To suppress this, the sky should be measured repeatedly with a scanning rate faster than the typical value of $f_{knee}=0.1$ Hz \cite{Tajima2012}. The telescope scans the sky with continuous rotations at 20 rpm (revolutions per minute) and an elevation of 60 degrees. The rotation speed results the scanning rate of 0.33 Hz which means that the telescope scans a ring every 3 seconds. A sky area of $10^{-6}$ sr (steradian) is observed on average 1000 times per day assuming a pixelization which divides the whole sky into 12,582,912 pixels.

\begin{figure}
    \begin{subfigure}{.5\textwidth}
        \centering
        \includegraphics[width=0.9\linewidth]{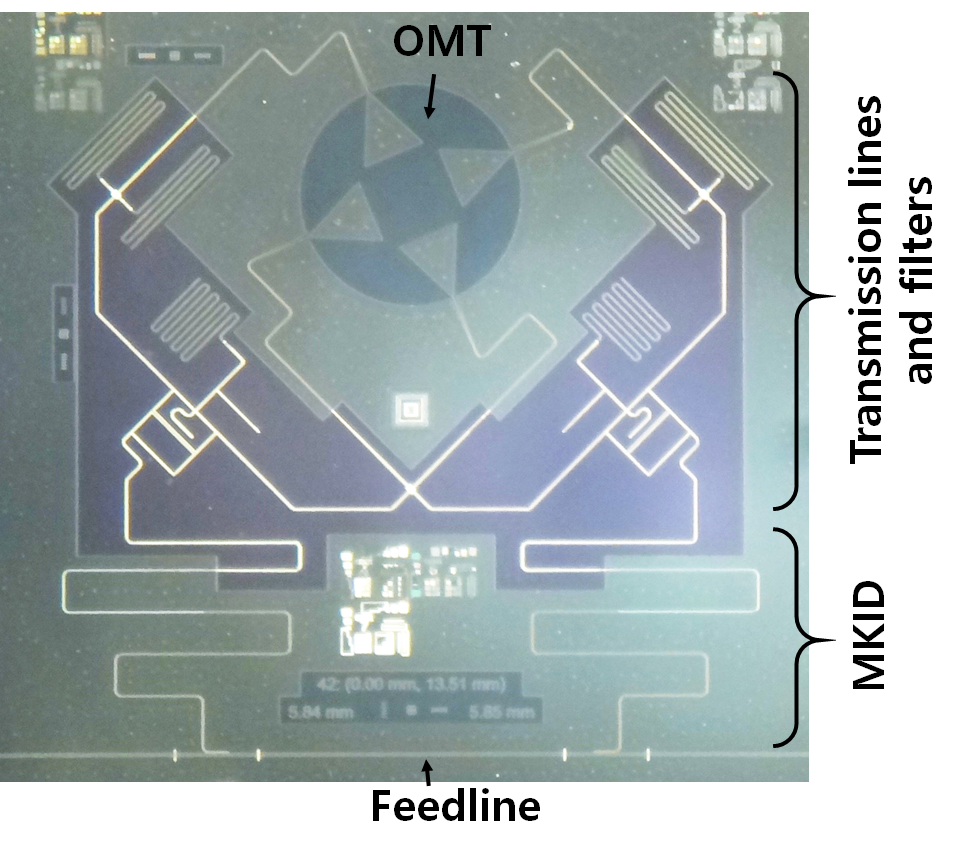}
        \caption{A pixel on a fabricated wafer.}
        \label{fig:MKID_pixel}
    \end{subfigure}
    \begin{subfigure}{.5\textwidth}
        \centering
        \includegraphics[width=.9\linewidth]{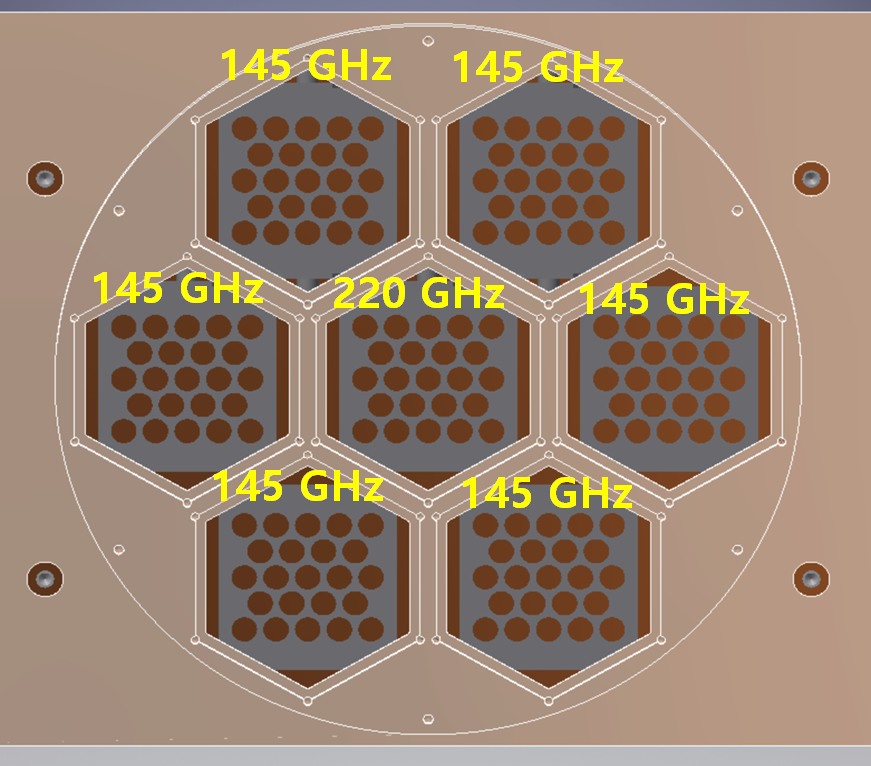}
        \caption{Focal plane design of GroundBIRD.}
        \label{fig:MKID_design}
    \end{subfigure}
    \caption{The Focal plane of GroundBIRD. A pixel on a fabricated wafer and the design of entire focal plane with 7 modules are shown in (a) and (b), respectively.}
    \label{fig:MKID_main}
\end{figure}

The MKID is a photon sensor using superconducting resonators \cite{Day}. An MKID array for the focal plane is under fabrication in RIKEN (Institute of Physical and Chemical Research). 
Figure \ref{fig:MKID_pixel} shows a full scale design of our MKID. An orthomode transducer (OMT) in each pixel receives photons in two perpendicular directions. The signals from the OMTs are transmitted to the MKIDs through the identical circuits with transmission line converters and filters, and the MKIDs detect the intensities of the photon in two directions.
The meander line at the end of the circuit is MKID as shown in Fig. \ref{fig:MKID_pixel}. The resonant frequencies for our MKIDs are controlled by their length, and are in the range of 4 to 8 GHz. The OMTs will be coupled with a corrugated horn antennae or a hemispherical lenslets for higher optical efficiency. Due to the fast response and short dead time of MKID, our system can have a fast sampling rate of 1 kSps.

The focal plane of the GroundBIRD telescope has 7 modules as shown in Fig. \ref{fig:MKID_design}. GroundBIRD will cover two frequency bands, 145 GHz and 220 GHz, for the separation of dust foreground. 
The module in the center is for 220 GHz and outer 6 modules are for 145 GHz. 
Each module has 23 pixels and the total number of pixels are 161 of which 138 for 145 GHz and 23 for 220 GHz. The sensitivity for the sky signal is determined by the integration time for observation per pixel. The expected sky pixel noise level is 21 $\mu$K arcmin by assuming three-year observation, 138 pixels and sky noise equivalent temperature (NET) of 300 $\mu$K. By combining with the large sky coverage, this value guarantees that the GroundBIRD can measure the tensor-to-scalar ratio from B-mode polarizations patterns at large angular scales ($l \sim 6$).

\begin{figure}
    \begin{subfigure}{0.5\textwidth}
        \centering
        \includegraphics[scale=0.45]{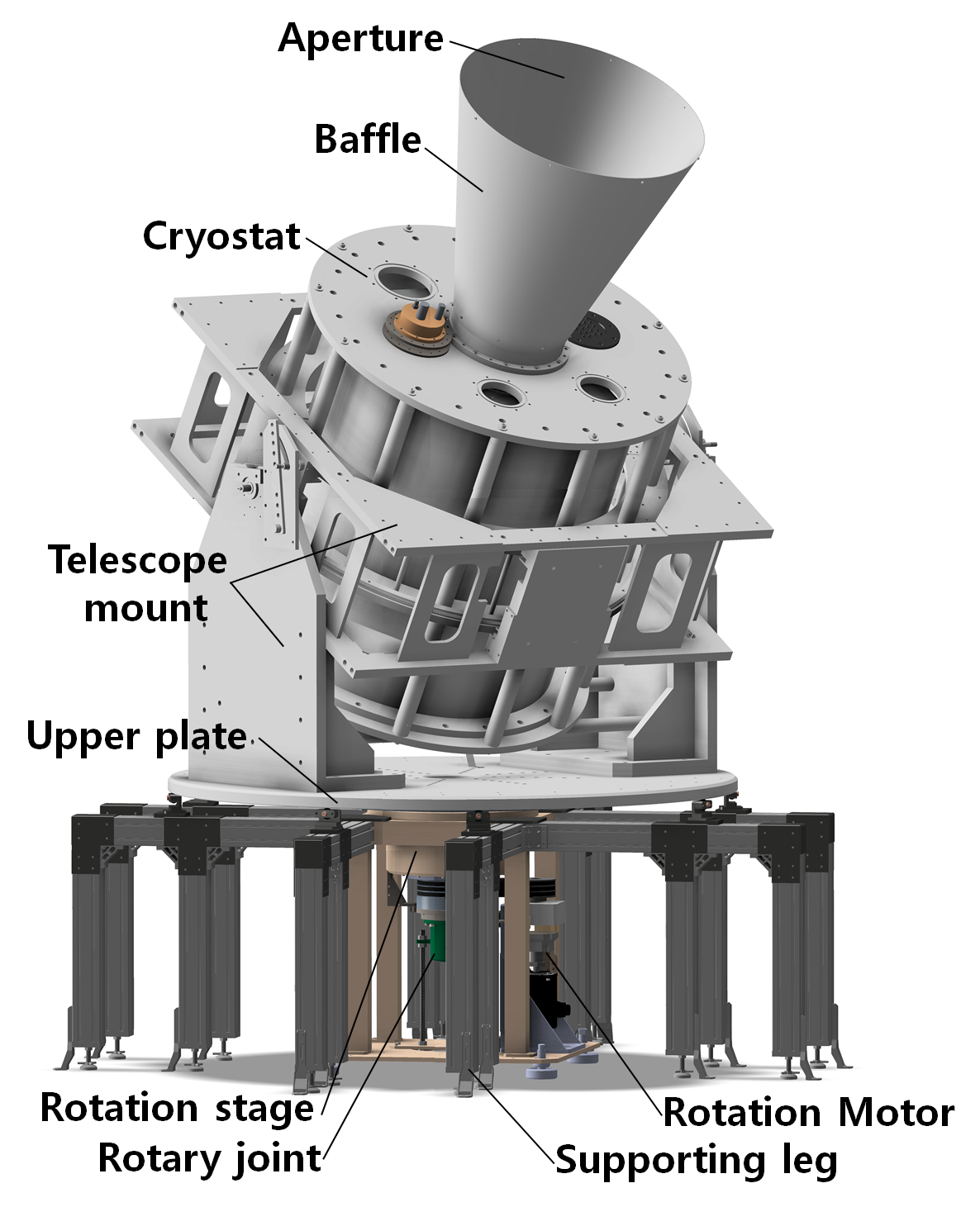}
        \caption{The GroundBIRD telescope.}
        \label{fig:GB_entire}
    \end{subfigure}
    \begin{subfigure}{0.5\textwidth}
        \centering
        \includegraphics[scale=0.95]{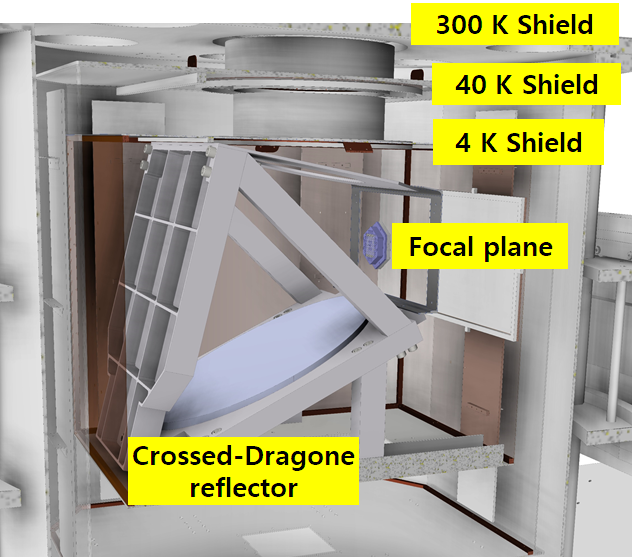}
        \caption{Inside the cryostat}
        \label{fig:GB_inside}
    \end{subfigure}
    \label{fig:}
    \caption{3D modeling of GroundBIRD. The entire telescope is shown in (a), and (b) shows inside the cryostat.}
\end{figure}

The cryostat of the GroundBIRD telescope is a cylindrical vacuum chamber with diameter of 1.5 m and 1.5 m in height as shown in Fig. \ref{fig:GB_entire}. Within the chamber, there are 40 K and 4 K stages. The temperatures of theses stages are maintained by two-stage pulse tube cooler (PTC). 
The 4 K stage contains the main part of the telescope including optics and focal plane. It is insulated by 40 K and 4 K shields. The shields are covered with multilayer insulation (MLI) sheets. In addition, these are wrapped in anti-magnetic sheets to reduce the effect of external magnetic fields on the MKID array \cite{Kutsuma2018}.
The aperture windows of 40 K and 4 K shields are covered by low-pass metal mesh filters \cite{Metalmesh} and radio transparent multilayer insulator (RT-MLI) \cite{Choi2013}. 

The crossed-Dragone reflector is the main optics of the GroundBIRD telescope \cite{Tajima2012}. The incident radiations are reflected on the primary and secondary mirrors. The two reflectors focus a plane wave to a point on the focal plane. The resulting beam width for a pixel in the sky is about 0.5 degrees by this focusing. The crossed-Dragone reflector is cooled down on the the 4 K stage. By cooling the reflector, thermal noise caused by the surface emission can be suppressed \cite{Tajima2012}. 

The telescope is rotated by our rotation table. It consists of the rotation stage in the center, supporting legs and the upper plate. The rotation stage rotates the upper structure. A custom-made rotary joint is used for the electricity supply, and helium gas circulation during the rotation \cite{Tajima2012}.

The observation site is Teide Observatory in Tenerife, Canary islands, Spain. The altitude and longitude are $28\degree 18'$N and $16\degree 30'$W and the altitude is 2390 m.

Even for only one day observation, we can make a sky map of a strip centered at declination of $28\degree$ with width of $60\degree$. The resulting coverage is 40\% of the full sky and angular resolution is 0.5 degrees. 
To improve the signal-to-noise ratio, we will stack the maps for year-scale. We expect to obtain the angular power spectrum for the angular scales of $6 < l < 300$, which enables to observe both recombination ($l \sim 100$) and reionization ($l < 10$) bumps in B-mode spectrum \cite{Tajima2012}. 

\section{Status of GroundBIRD}

In the first half of this year, the GroundBIRD telescope has been transported from Japan to IAC in Tenerife, Spain. 
We are assembling the telescope and preparing its deployment on the Teide observatory.
We shipped our telescope after we confirmed the receiver performances with rotation on the telescope mount. The temperature of focal plane and cold optics during rotation at 20 rpm. For 16 hours after the start of rotation, the cold optics temperature is maintained at 3.5 K and the temperature of the focal plane is maintained between 250 and 270 mK. Since the detector temperature varies much slower than the rotation of the telescope, the drift in detector response correlated with the temperature change is suppressed with the $1/f$ noise removal algorithm during the data analysis. Additionally the resonant frequencies of MKIDs are calibrated every hour to compensate the shift of the resonance peak due to the temperature change.


\subsection{Detectors}

\begin{figure}
    \begin{subfigure}{.48\textwidth}
        \centering
        \includegraphics[width=0.8\linewidth]{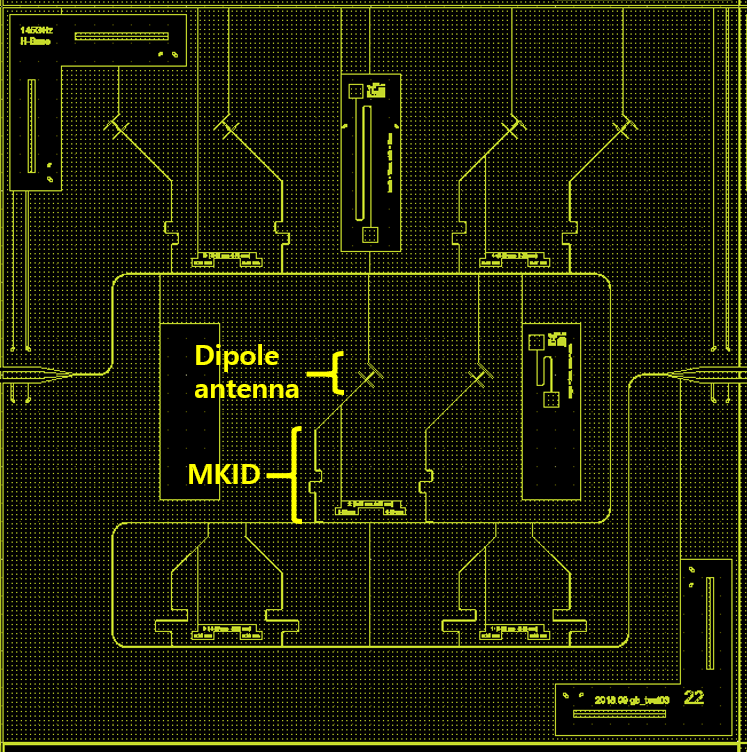}
        \caption{Mask design of test version MKID for 145 GHz. There are 10 MKIDs, each connected with a single polarized dipole antenna.}
        \label{fig:testver_MKID1}
    \end{subfigure}\hfill
    \begin{subfigure}{.48\textwidth}
        \centering
        \includegraphics[width=0.75\linewidth]{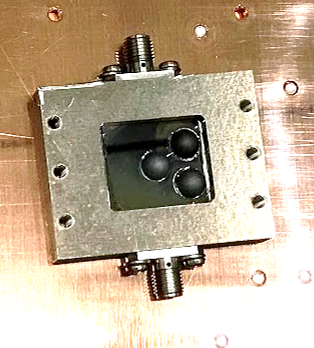}
        \caption{Fabricated test version MKID with lenslets. The hemispheres are the lenslets. Three pixels are coupled with the lenslets. It is mounted on the module for the focal plane.}
        \label{fig:testver_MKID2}
    \end{subfigure}
    \caption{Test version MKID. The mask design and fabricated wafer and lenslet are shown in (a) and (b) respectively.}
    \label{fig:testver_MKID}
\end{figure}

We fabricated a simplified version of MKID for the test. The mask design of the test version MKID is shown in \ref{fig:testver_MKID1}. The resonant frequencies are designed between 4 to 8 GHz. Each pixel has a single polarization antenna, coupled with a lenslet which is shown in Fig. \ref{fig:testver_MKID2}. The hemispherical lenslet diameter is 6 mm, which determines the spacing between the detectors.

\begin{figure}[!htp]
    \begin{subfigure}{0.48\textwidth}
        \centering
        \includegraphics[width=1\linewidth]{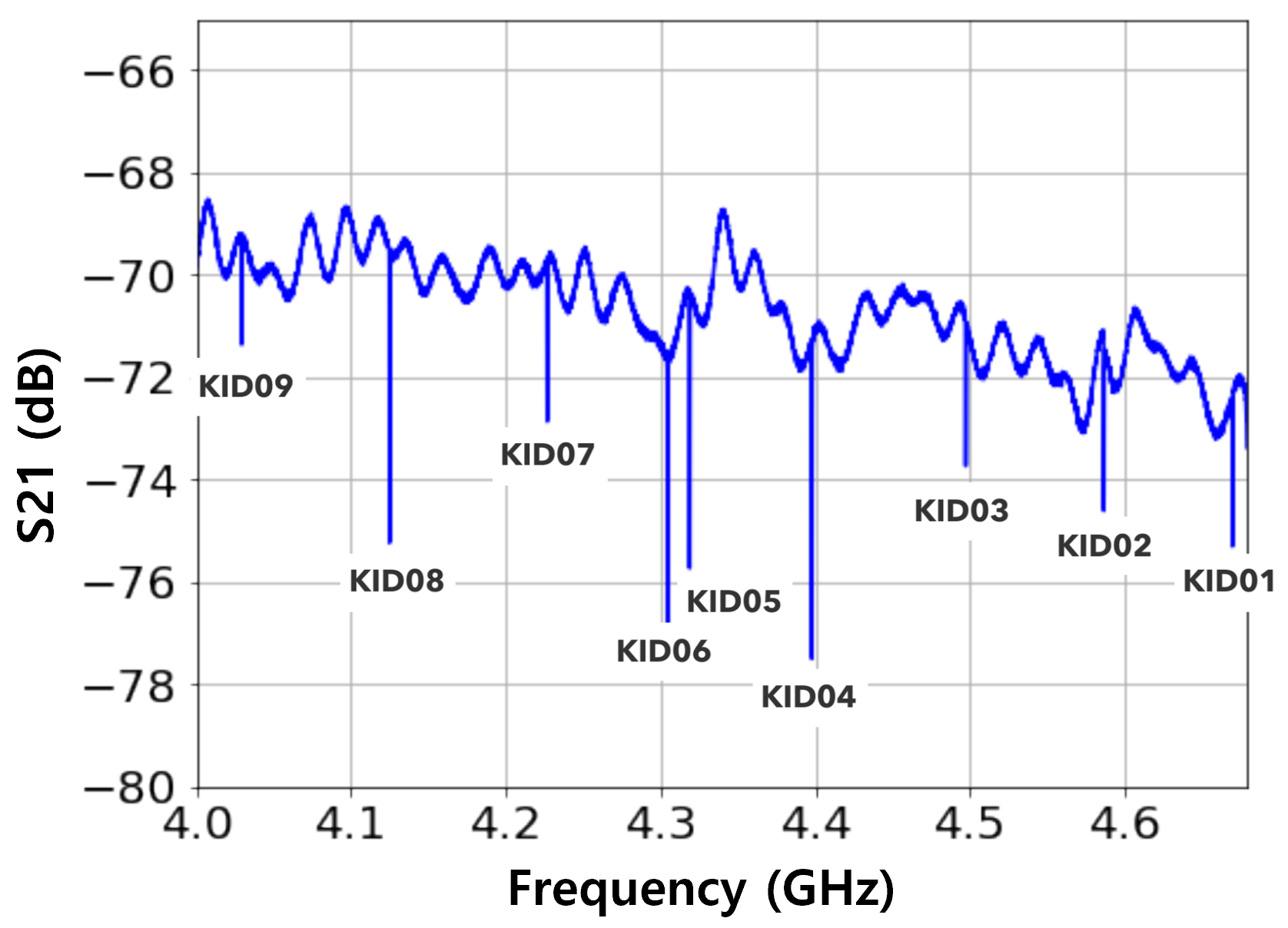}
        \caption{S21 of the test version MKID. The 9 peaks are observed between 4.0 to 4.7 GHz. The peaks are spaced approximately 100 MHz from each other except for the KID05.}
        \label{fig:resonances}
    \end{subfigure}\hfill
    \begin{subfigure}{0.48\textwidth}
        \centering
        \includegraphics[width=1\linewidth]{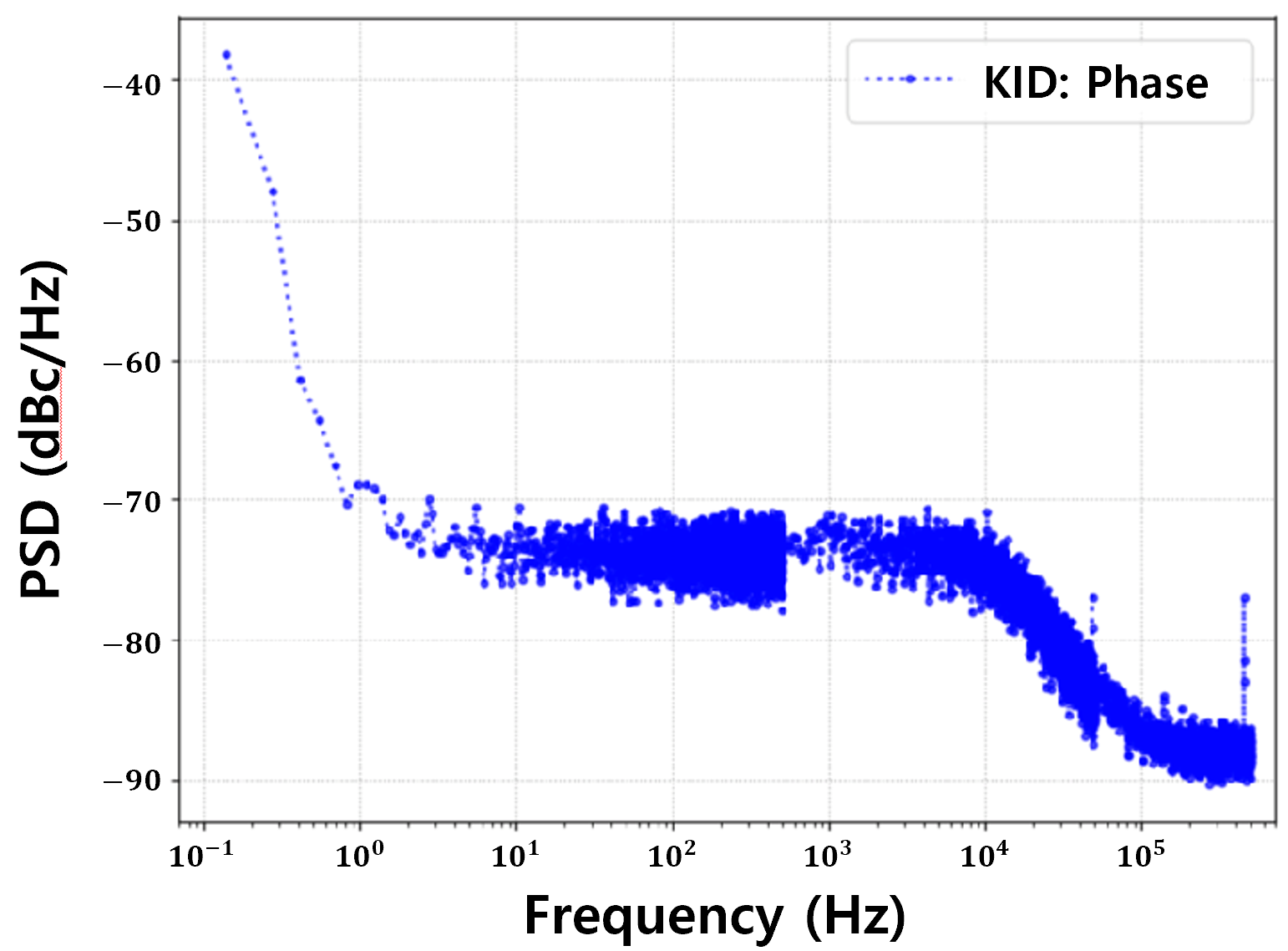}
        \caption{The phase noise PSD of an MKID. The knee frequency is 1 Hz and the roll-off frequency is $10^4$ Hz. The flat noise psd level is  }
        \label{fig:psd}
    \end{subfigure}
    \caption{Lab tests of the test version MKID. The S21, measured with VNA, and the phase noise PSD for an MKID are shown in (a) and (b) respectively.}
    \label{fig:kidmeas}
\end{figure}

The lab test of the test version MKID are shown in Fig. \ref{fig:kidmeas}. The amplitude of S21 is measured with a vector network analyzer (VNA). The 9 resonances are observed between 4 and 4.7 GHz where one missing peak is just outside this range. The resonance peaks are spaced approximately 100 MHz from each other except for one peak. The phase noise power spectral density (PSD) of an MKID is shown in Fig. \ref{fig:psd}. The knee frequency is 1 Hz which is below the scanning rate of 3 Hz.

\begin{figure}
    \begin{subfigure}{.6\textwidth}
        \centering
        \includegraphics[width=.9\linewidth]{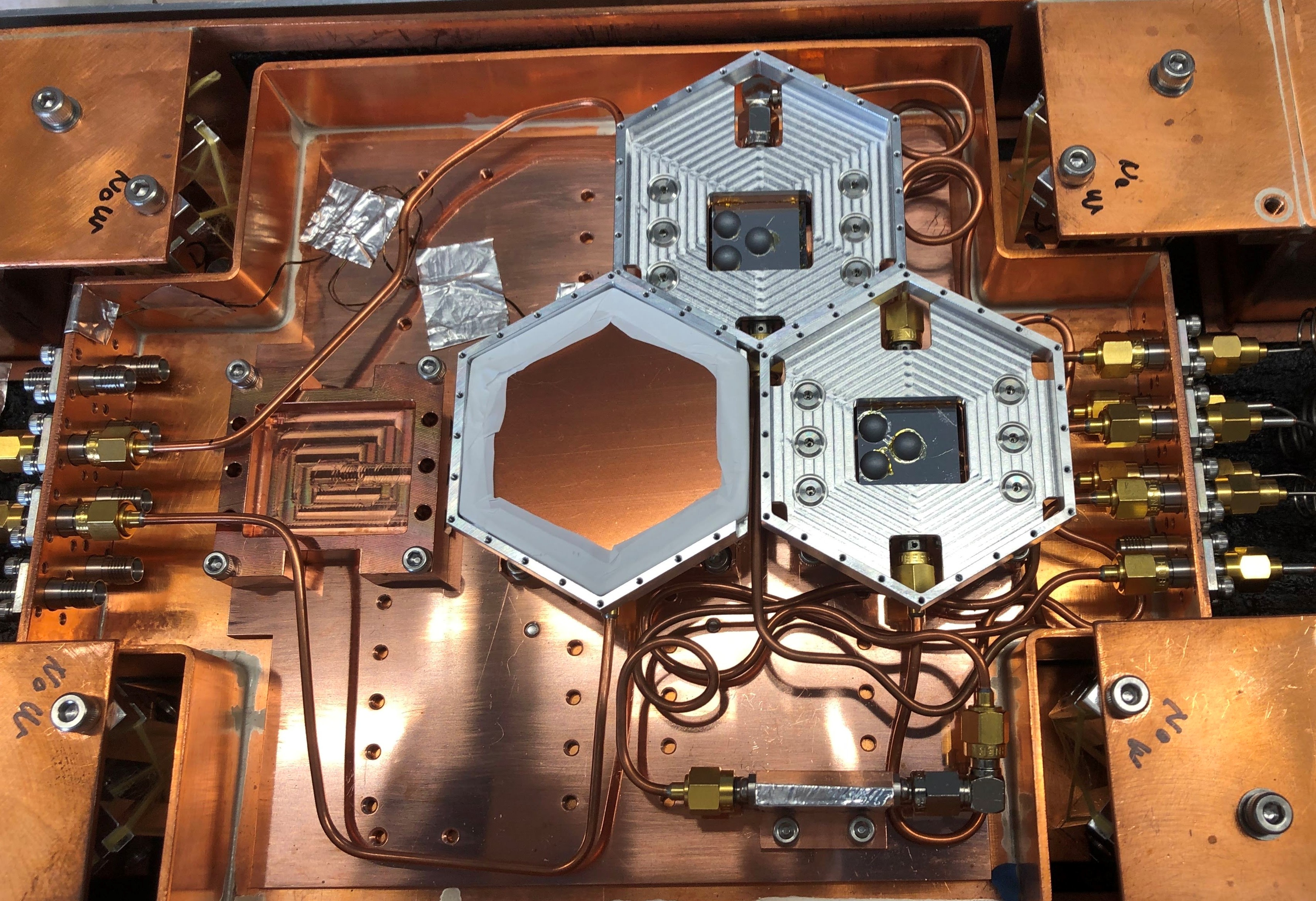}
        \caption{MKID arrays on the focal plane. Three MKID arrays are installed. The plate on the center array is a metal mesh filter.} 
        \label{fig:focal_plane_setup}
    \end{subfigure}
    \begin{subfigure}{.4\textwidth}
        \centering
        \includegraphics[width=.8\linewidth]{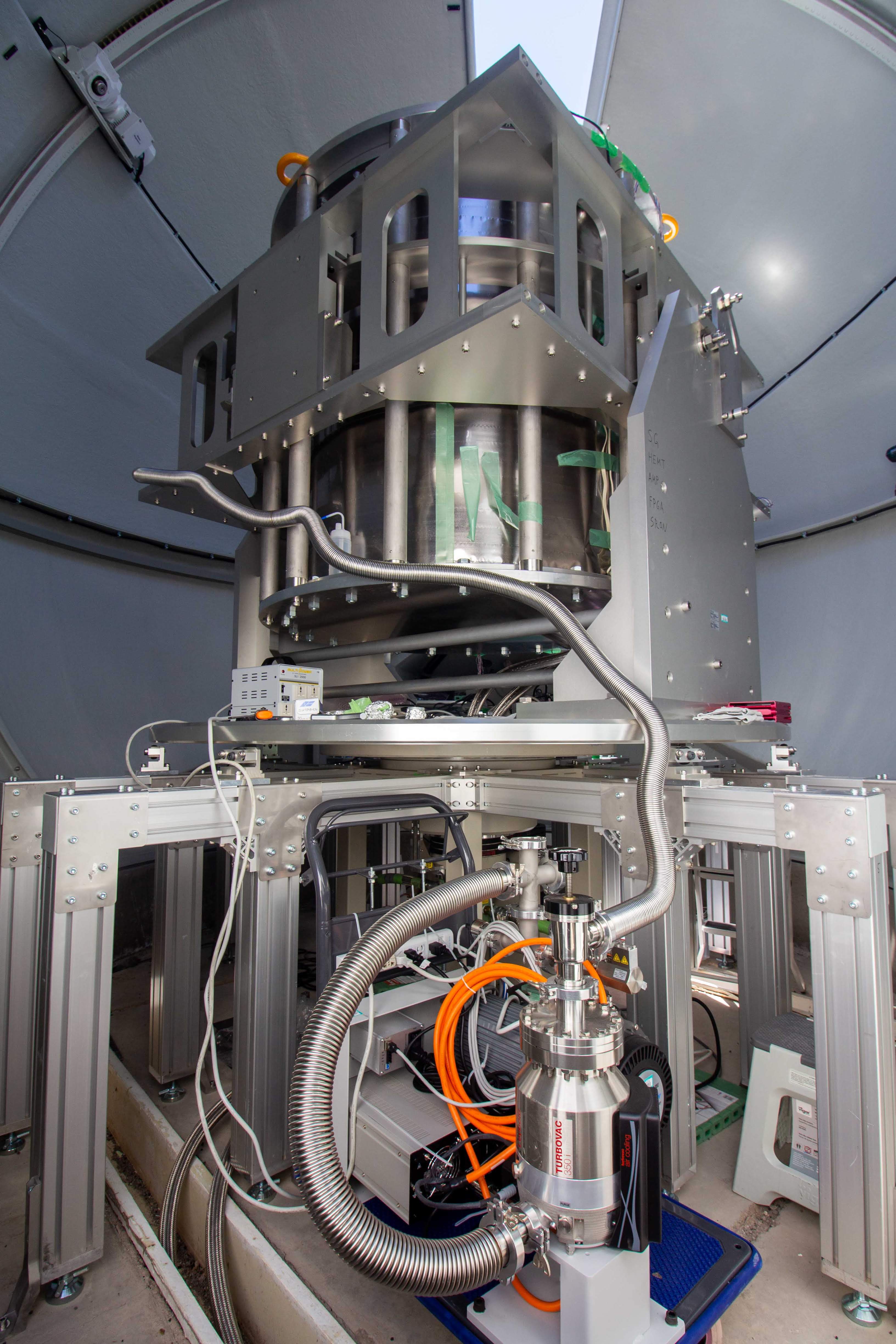}
        \caption{Installed telescope in the dome.}
        \label{fig:installed_telescope}
    \end{subfigure}
    \caption{The current setup of the focal plane and the installed telescope.} 
    \label{fig:installed_instruments}
\end{figure}

The MKID arrays are set up in the telescope as shown in Fig. \ref{fig:focal_plane_setup} and will be used for the first light and in early stages of the commissioning. 

\subsection{Telescope installation and commissioning}

A Very Small Array (VSA) enclosure is used for the GroundBIRD experiment. A dome built by baader planetarium company \cite{baader} is placed at the site and the rotation table was installed inside the dome. The telescope was mounted on the rotation table as shown in Fig. \ref{fig:installed_telescope} and the installation is finished.  
The commissioning for calibrations of the detector and pointing of the telescope is being performed by observing the moon and the tau A. 

\section{Conclusions}

GroundBIRD aims to observe the CMB B-mode polarization and measure the tensor-to-scalar ratio $r$ with a precision of $\sigma_r \sim 0.01$.
The GroundBIRD telescope was moved from Japan to Tenerife and installed at the Teide observatory. The commissioning of the telescope is on going and the MKID array for the actual observation is under development aiming to start observation in 2020.

\begin{acknowledgements}

This work was supported by JSPS KAKENHI Grant Numbers JP15H05743, JP16J09435, JP18H05539, JP18J01831, JP15K13491, JP19H01916, JPR2804, and by the NRF Grant Number NRF-2017R1A2B3001968. This was also supported by Kyoto University and MEXT SPIRITS grant. We also thank Hisao Kawano, Noboru Furukawa, Hiroshi Watanabe, Advanced Technology Center of National Astronomical Observatory of Japan, and Advanced Manufacturing Support Team of RIKEN.

\end{acknowledgements}


\end{document}